\documentclass{nic-series}

\begin{document} 

\title{First-Principles Many-Body Investigation of Correlated Oxide Heterostructures:
Few-Layer-Doped SmTiO$_3$}

\author{Frank Lechermann}

\institute{I. Institut f{\"u}r Theoretische Physik, Universit{\"a}t Hamburg,\\
           D-20355 Hamburg, Germany\\
           \email{Frank.Lechermann@physnet.uni-hamburg.de}
           \and
           Institut f{\"u}r Keramische Hochleistungswerkstoffe, Technische Universit\"at
           Hamburg-Harburg,\\ 
           D-21073 Hamburg, Germany}

\maketitle

\begin{abstracts}
Correlated oxide heterostructures pose a challenging problem in condensed matter research due
to their structural complexity interweaved with demanding electron states beyond the 
effective single-particle picture. By exploring the correlated electronic structure of
SmTiO$_3$ doped with few layers of SrO, we provide an insight into the complexity of such
systems. Furthermore, it is shown how the advanced combination of band theory on the level
of Kohn-Sham density functional theory with explicit many-body theory on the level of dynamical
mean-field theory provides an adequate tool to cope with the problem. Coexistence of 
band-insulating, metallic and Mott-critical electronic regions is revealed in individual 
heterostructures with multi-orbital manifolds. Intriguing orbital polarizations, that 
qualitatively vary between the metallic and the Mott layers are also encountered. 
\end{abstracts}

\section{Introduction}
Research on oxide heterostructures emerged in the beginning of the 2000s as a novel topical
field and belongs nowadays to a key focus in condensed matter and materials science (see e.g.
Refs.~\citen{hwa12,cha14,jan16} for reviews). Thanks to 
important advancements in experimental preparation techniques, the design of 
oxide materials, e.g. by joint layering of different bulk compounds, opens new possibilities
to devise matter beyond nature's original conception. Importantly, oxide heterostructures
are not only relevant because of their potentially future technological importance, but they
also challenge known paradigms in condensed matter physics. For instance, the obvious traditional 
separation of electronic materials into metals, band insulators or Mott insulators may be
reconsidered in such materials. Since known bulk features of individual oxide 
building blocks disperse within a given heterostructure, characterisics of various electronic
signatures may be detected~\cite{lec15,lec17}. Notably, the distinguished role of interface 
physics within a demanding quantum-mechanical environment is one of main concerns in this area.
\newline
Density functional theory (DFT) in the Kohn-Sham representation is the standard tool for
materials science starting from the atomic scale. However, this theoretical approach has its
flaws for systems where the mutual interaction among the condensed matter electrons is
comparable or even larger than the dispersion energy from hopping on the underlying lattice. 
For various reasons, many interesting oxide heterostructures are located in the latter regime, 
marking them as correlated oxide heterostructures (COHs). Modeling these fascinating designed
materials, including the possibility for further engineering and prediction of intriguing
phenomenology, therefore asks for a theoretical approach beyond effective single-particle
theory. The combination of DFT with the explicit many-body framework of dynamical mean-field
theory (DMFT), the so-called DFT+DMFT method, represents such an 
approach (see e.g. Ref.~\citen{kotliar_review} for a review).
\newline
In the present work, the charge self-consistent DFT+DMFT framework is put into practice to
examine the correlated electronic structure of Mott-insulating SmTiO$_3$ doped with a few
layers of SrO in an heterostructure architecture. This illustrates the challenges of COHs
as well as the capabilities of advanced electronic structure theory to address those.
Coexistence of different electronic phases is encountered, ranging from Mott-insulating,
metallic up to band-insulating.

\section{SrO doping layers in SmTiO$_3$}
\label{sec:srolayers}
\begin{figure}[b]
\begin{center}
\includegraphics*[width=11.5cm]{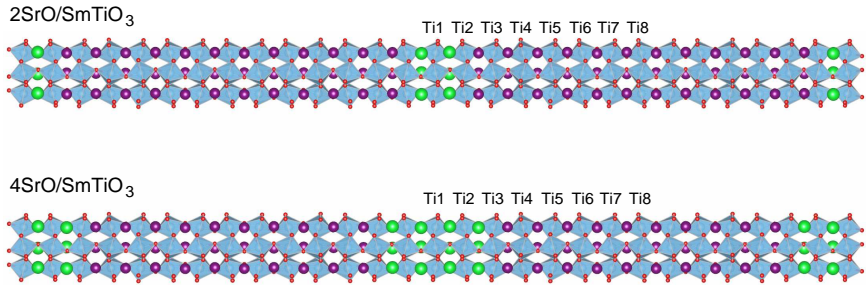}
\caption{\label{fig:layers}
Supercells for the cases of 2- and 4SrO-layer doping of SmTiO$_3$ with the horizontal axis
as the stacking direction (the $n=6$-layer system is constructed correspondingly). 
Sr: green, O: red, Sm: violet, Ti: blue. There 
are eight inequivalent Ti sites Ti1-8, describing the layer dependence of the electronic 
structure.
Each layer consists of two Ti sites in the unit cell in order to incorportate the relevant
orthorhombic distortions, but those intra-layer sites are are treated equivalent by 
symmetry.
}
\end{center}
\end{figure}
The rare-earth titanate SmTiO$_3$ is a distorted perovskite with orthorhombic $Pbnm$ 
crystal-symmetry group. Electronically, it is a Mott insulator in the bulk~\cite{kom07}, i.e. 
electrons are localized in real space and cannot metallize the compound via hopping because of 
the strong Coulomb repulsion. For this system, the most-relevant Coulomb impact is on the 
Ti-$3d$ shell, which is nominally filled with one electron since titanium is in a 
formal Ti$^{3+}$ state. Moreover, calculations show that the electron dominantly
resides in a single effective orbital of $t_{2g}$ kind~\cite{lec17}, i.e. orbital polarization
is an important issue.
\newline
In the present work, a well-defined doped Mott-insulator shall be investigated in a 
heterostructure architecture by inserting $n=2,4,6$ layers of SrO into SmTiO$_3$, 
thereby replacing SmO layers, respectively (see Fig.~\ref{fig:layers}). Because of the 
different valence of Sr$^{2+}$ compared to Sm$^{3+}$, this leads to an effective hole doping. 
Experimental transport studies of such systems have recently been performed~\cite{jac14,mik15}.
To model these complex COHs in a first-principles setting, superlattices based on 140-atom 
unit cells are here considered. They consist of 14 TiO$_2$ layers, and each layer is build 
from two symmetry-equivalent-treated Ti ions. With the inserted SrO layers, there are then 
8 Ti sites different by symmetry, located in different layers (cf. Fig.~\ref{fig:layers}). 
The original lattice parameters~\cite{kom07} are brought in the directional 
form of the experimental works~\cite{jac14,mik15}, but without lowering the 
$Pbnm$ symmetry. The original $c$-axis is {\sl parallel} to the doping layer and the original 
$a,b$-axes are respectively inclined. 
With fixed lattice parameters, all atomic positions in the supercell are structurally 
relaxed within DFT(GGA) until the maximum individual atomic force settles below 5\,mRyd/a.u.. 
The lattice distortions introduced by the inserted SrO layers are well captured by this
approach. 
\newline
Note that the $\delta$-doped case of $n=1$ has been recently studied in detail by a similar
approach~\cite{lec17}. In this respect, the present work advances on this previous study and
renders it possible to follow the evolution of the correlated electronic structure of 
heterostructure-doped SmTiO$_3$ with further SrO layers. 

\section{Theoretical Approach}
The charge self-consistent DFT+DMFT method combines band theory and many-body theory on
an equal footing. The band-theoretical aspect is delivered on the DFT level, and through
a downfolding to a correlated subspace of relevant sites and orbitals, electronic 
correlations are evaluated in the many-body scope. Those correlations define an electronic
self-energy that reenters the DFT level in updating the Kohn-Sham potential. Thereby
a self-consistency cycle is defined that at convergence provides the many-body electronic 
structure beyond conventional exchange-correlation functionals~\cite{sav01,pou07,gri12}.
For the DFT part, a mixed-basis pseudopotential coding~\cite{mbpp_code}, based on 
norm-conserving pseudopotentials as well as a combined basis of localized functions and plane 
waves, is utilized. We here employ the generalized-gradient approximation (GGA) in the 
Perdew-Burke-Ernzerhof form~\cite{per96}.
\newline
Here, we aim for a description of SrO layers in SmTiO$_3$. 
The three Ti-$3d(t_{2g})$ orbitals, split-off from the remaing $e_g$ orbitals of the full 
$3d$ shell, host the key-relevant single electron of SmTiO$_3$.
The correlated subspace therefore consists of the effective Ti($t_{2g}$) Wannier-like 
functions, i.e. is locally threefold. These functions are obtained 
from a projected-local-orbital formalism~\cite{ama08,ani05}, using as projection 
functions the linear combinations of atomic $t_{2g}$ orbitals. The latter diagonalize the Ti  
orbital-density matrix from DFT. A band manifold of 60 $t_{2g}$-dominated 
Kohn-Sham states at lower energy are used to realize the projection. Local Coulomb 
interactions within the correlated subspace in the so-called Slater-Kanamori form of
a multi-orbtial Hubbard Hamiltonian are parametrized by a Hubbard $U=5\,$eV and a 
Hund's coupling $J_{\rm H}=0.64\,$eV~\cite{pav05}. The 8 coupled single-site DMFT impurity 
problems in the supercells are, respectively, solved by the continuous-time quantum 
Monte Carlo (QMC) scheme~\cite{rub05,wer06} as implemented in the TRIQS 
package~\cite{par15,set16}. 
A double-counting correction of fully-localized type~\cite{ani93} is utilized, which 
accounts for correlation effects already included on the GGA level. About 40-50 DFT+DMFT 
iterations (of alternating Kohn-Sham and DMFT impurity steps) are necessary for full 
convergence. Note that DFT+DMFT, contrary to conventional DFT, explicitly treats finite
temperature. In all calculations presented in this scope, the temperature was set to
$T=145$\,K.
\newline
The large-scale calculations run in a parallelized computing architecture for the $k$-points
of the DFT part as well as for the QMC sweeps. Computations also ask for a sizable memory due 
to the demanding supercell structures. Even on supercomputing machines, full convergence for
an individual superlattice at a given finite temperature still asks for several days of 
computing time.

\section{DFT+DMFT results for few-layer-doped SmTiO$_3$}
\begin{figure}[b]
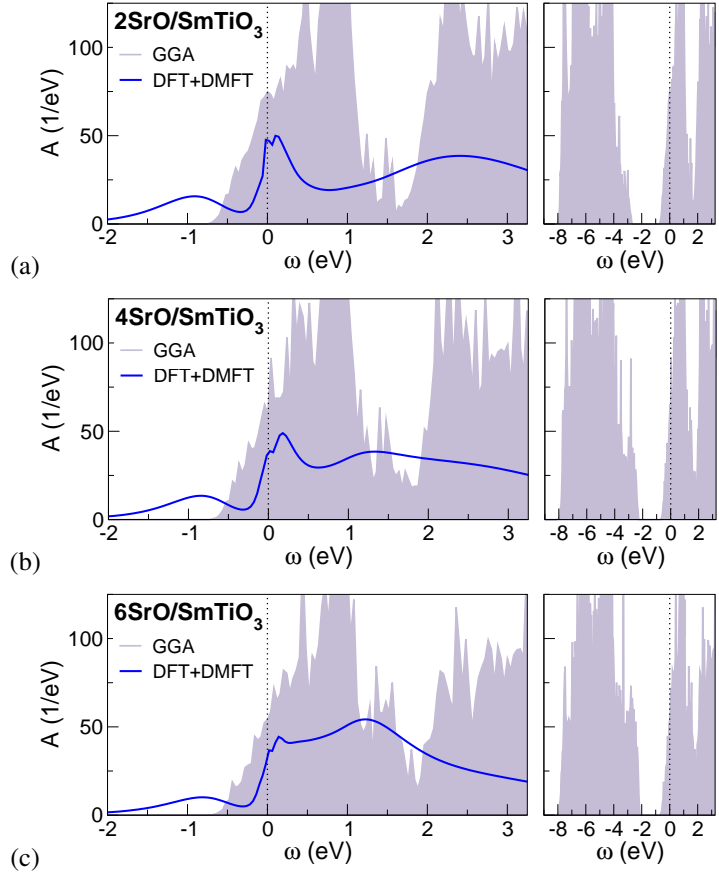

\begin{center}
(a)\includegraphics*[width=9cm]{ldadmft-dos.2smto.eps}\\[0.2cm]
(b)\includegraphics*[width=9cm]{ldadmft-dos.4smto.eps}\\[0.2cm]
(c)\includegraphics*[width=9cm]{ldadmft-dos.6smto.eps}
\caption{\label{fig:totspec}
Total spectral function of SmTiO$_3$ doped with two (a), four (b) and six (c) layers
of SrO compared to DFT(GGA). Right part shows the DFT density of states in a larger
energy window, respectively, displaying the dominantly O-$(2p)$ block for energies
$\sim [-8,-2]$\,eV, the $t_{2g}$-like part around the Fermi level and the 
$e_g$-dominated states above 2\,eV.}
\end{center}
\end{figure}
Key interest is in both, the global behavior as well as the layer-resolved physics of these
artifical $n$SrO/SmTiO$_3$ systems. Importantly, the given structurally well-defined doping
of the Mott insulator, enables an account of the realistic many-body electron states which
is free of the usual disorder effects in common bulk-doped materials.
\newline
We first focus on the global electronic system by inspecting the total spectral function,
plotted in Fig.~\ref{fig:totspec}. In the effective single-particle picture of conventional
DFT this function coincides with the density of states (DOS). The relevant DOS building  
blocks for the present transition-metal oxides are a dominantly O-$2p$ spectral part deep
in the occupied region, a Ti-$3d(t_{2g})$-like part at low energy and a Ti-$3d(e_{g})$-like
part energetically higher in the unoccupied region. 
As a first observation, both theoretical schemes, DFT(GGA) and DFT+DMFT, mark the present 
COHs as metallic, in line with experiment~\cite{jac14,mik15}. In addition, the $n=1$ case of 
$\delta$-doped SmTiO$_3$ is identified as conducting~\cite{jac14,mik15,lec17}.
\begin{table}[b]
\centerline{
\begin{tabular}{c|c|rrrrrrrr}
SrO layers $n$ & orbital & Ti1  & Ti2  & Ti3  & Ti4  & Ti5 & Ti6 & Ti7 & Ti8 \\ \hline
           & $|1\rangle$  & 0.11 & 0.13 & 0.24 & 0.39 & 0.07 & 0.05 & 0.21 & 0.04 \\
2          & $|2\rangle$  & 0.13 & 0.13 & 0.53 & 0.46 & 0.89 & 0.93 & 0.71 & 0.96 \\
           & $|3\rangle$  & 0.04 & 0.26 & 0.14 & 0.13 & 0.02 & 0.02 & 0.04 & 0.01 \\[0.15cm]
           & $|1\rangle$  & 0.05 & 0.05 & 0.09 & 0.34 & 0.24 & 0.04 & 0.18 & 0.03 \\
4          & $|2\rangle$  & 0.04 & 0.04 & 0.08 & 0.35 & 0.63 & 0.93 & 0.81 & 0.96 \\
           & $|3\rangle$  & 0.02 & 0.05 & 0.31 & 0.18 & 0.08 & 0.01 & 0.02 & 0.01 \\[0.15cm]
           & $|1\rangle$  & 0.03 & 0.03 & 0.06 & 0.09 & 0.39 & 0.06 & 0.16 & 0.03 \\
6          & $|2\rangle$  & 0.02 & 0.02 & 0.05 & 0.09 & 0.32 & 0.90 & 0.82 & 0.97 \\
           & $|3\rangle$  & 0.03 & 0.02 & 0.06 & 0.21 & 0.13 & 0.02 & 0.01 & 0.01 
\end{tabular}
}
\caption{Ti($t_{2g}$) occupations from DFT+DMFT in the crystal-field basis within each 
TiO$_2$ layer of $n$-layer-doped SmTiO$_3$.}\label{tab:occ}
\end{table}
\newline
For a comparison of the different $n$SrO-doping cases it is important to realize that 
replacing $n>1$ SmO layers with SrO ones actually results in new finite building blocks of 
effective SrTiO$_3$. In bulk form, the latter perovskite is a band insulator with nominal
Ti-$3d^0$ filling. Thus by increasing the number of SrO layers, the originally Mott-insulating
system is replaced in parts by a band-insulating system. It is a particularly interesting
scenario to have the different insulator concepts, i.e. the band insulator from band theory
and the Mott insulator from interacting many-body theory, conjoint within a single electronic
structure problem. Concretely, this means that the present doping with more and more SrO 
layers should not simply result in a successive strengthening of the metallic character.
There are 28 Ti atoms in the supercell, which from $3d^1$ yields also 28 electrons in the 
occupied part of the spectral function. Each SrO layer, incorporating two Sr sites, adds two
holes, resulting in a nominal doping of 1/14 holes per Ti site. Accordingly, from $n=2$ to
$n=6$ the Ti-dominated occupied part of the spectrum shrinks from 28 to 16 electrons, as 
visualized in Fig.~\ref{fig:totspec}. The degree of metallicity is more elusive since to a first 
approximation encoded in the height and width of the quasiparticle (QP) peak around the
Fermi level. For $n=2$, the QP peak is higher than for $n=4$, but the width is slighly larger
for the latter case. Compared to these cases, the QP peak is clearly diminished for $n=6$.
At higher energies in the occupied spectrum, DFT+DMFT accounts for the lower Hubbard band
at $\sim-0.9$\,eV, denoting the degree of real-space localization, which is missing in the 
DFT description. The transfer of spectral weight from the QP peak to the Hubbard peaks with
increasing correlation strength is a hallmark of strongly correlated systems.
\begin{figure}[t]
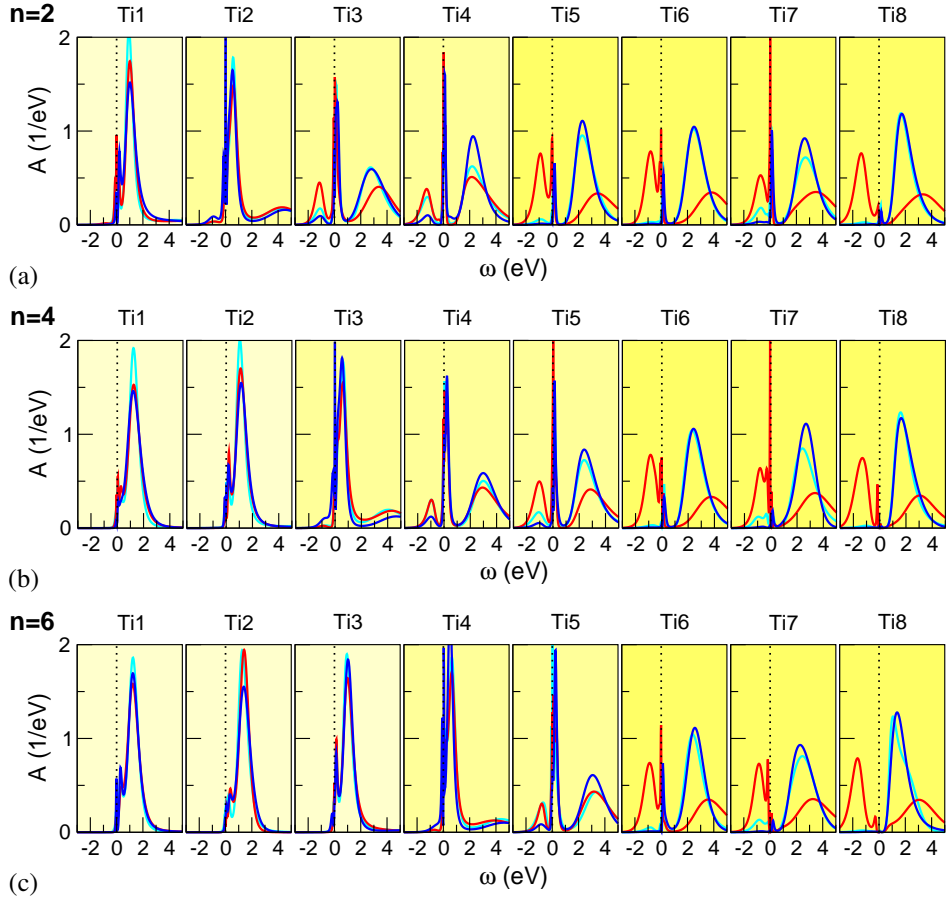

\begin{center}
(a)\hspace*{-0.4cm}\includegraphics*[width=12.5cm]{local2dmft.eps}\\[0.2cm]
(b)\hspace*{-0.4cm}\includegraphics*[width=12.5cm]{local4dmft.eps}\\[0.2cm]
(c)\hspace*{-0.4cm}\includegraphics*[width=12.5cm]{local6dmft.eps}
\caption{\label{fig:locspec}
Local Ti- and orbital-resolved DFT+DMFT spectral function of SmTiO$_3$ doped with two (a), 
four (b) and six (c) layers of SrO. Colored curves denote the orbital flavor:
$|1\rangle$: lightblue, $|2\rangle$: red and $|3\rangle$: blue.
Different colored backgrounds mark tendencies towards band-insulating (bright), metallic 
(light yellow) and Mott-critical (yellow) character.}
\end{center}
\end{figure}
\newline
The competition between band-insulating and Mott-insulating tendencies becomes more 
obvious on the layer- and orbital-resolved local level. In the orthorhombic distorted
systems, the Ti$(t_{2g})$ orbitals $|xz\rangle$, $|yz\rangle$ and $|xy\rangle$ hybridze 
on each site, forming effective crystal-field orbitals $|1\rangle$, $|2\rangle$ and 
$|3\rangle$~\cite{pav05,lec15,lec17}. Though the linear combinations are layer-dependent,
the qualitative character remains stable throughout the TiO$_2$ layers. In bulk
SmTiO$_3$, the $d^1$ electron majorly resides in the $|2\rangle$ orbital.
Table~\ref{tab:occ} provides the multi-orbital fillings for the symmetry-inequivalent
Ti1-8 sites in $n$-layer SrO/SmTiO$_3$. Figure~\ref{fig:locspec} shows furthermore the
local spectral functions for each choice of $n$ for the given Ti sites, i.e. from the doping 
layers towards the bulk-like region of SmTiO$_3$. Interestingly, when embedded by the SrO 
layers, the system quickly establishes band-insulating-like behavior. Meaning, the Ti states 
become completely depleted and the region gets inaccessible for electron transport. On 
the other hand, far away from the SrO layers the system is Mott critical, i.e. either is
in a doped-Mott or Mott-insulating regime. Inbetween these different insulating(-like) 
parts, a seemingly moderately-correlated metallic region of 2-3 TiO$_2$-layers width 
is established, respectively. This metallic region shifts correspondingly with 
increasing $n$ in the superlattices. Surpisingly for each $n$, in a single TiO$_2$ layer 
of these metallic parts there is orbital polarization towards state $|3\rangle$, which
is of dominant $|xy\rangle$, i.e. inplane, contribution. Thereby associated is high
QP peak of identical orbital flavor. Since the system is strongly $|2\rangle$-polarized
in the Mott-critical region similar to the bulk compound, this additional polarization 
is a unique heterostructure effect. Note that it is absent for the $\delta$-doped case
$n=1$~\cite{lec17}, and hence could be important for the stronger Fermi-liquid character
in the cases $n>1$~\cite{jac14}.

\subsection{Summary}
Large-scale first-principles many-body calculations based on the advanced DFT+DMFT framework 
are capable of addressing the challenging emerging physics of correlated oxide heterostructure 
on a realistic level.
For the case of few-layer doped SmTiO$_3$, the coexistence of different electronic phases,
i.e. band-insulating, metallic and Mott-critical are predicted on a multi-orbital level.
In addition, alternating orbital polarizations are revealed, that open further engineering
possibilities. In general, the richness of various competing real-space regions of different
electronic kind in a single equilibrium system should enable various technological  
applications.

\section*{Acknowledgments}
The author gratefully acknowledges the computing time granted by the John von Neumann Institute
for Computing (NIC) and provided on the supercomputer JURECA at the J\"ulich Supercomputing 
Centre (JSC) under project number hhh08.

\bibliography{bibextra}

\end{document}